T H E   P A P E R :
-------------------

IMPACT PROOFS FOR THE WORK-HARDENING NATURE OF LOW- AND HIGH-Tc SUPERCON-
DUCTIVITY. POSSIBLE FORECAST OF THE PARAMETERS OF SUPERCONDUCTORS THROUGH
                THE ALL-RANGE TEMPERATURE TESTS

                                 *)
    Valery P. Kisel,  Adkham A. Paiziev   and   Arkady D. Styrkas
   Institute of Solid State Physics, 142432 Chernogolovka, Russia
  *) Institute of Electronics, Uzbek Akademy of Sciences, Tashkent

ABSTRACT
--------

The remarkable finding of this work is the linear correlation between
the critical temperature of superconducting (SC) transition, $T_c$, and the
room-temperature half-width of angular correlation of positron annihila-
----------------
tion phonons (ACPAP), $G_o/2$, in the series of powder samples of high-Tc
YBCO(123) ceramic superconductors (SCs) with various deficiency of oxygen
content. This correlation points to the existence of fundamental physical
parameter of materials in all-temperature range,which is the crucial one
for their insulator-metal-superconductor (IMSC) transition. We attribute
this key-parameter to the rapid increase and then the steady (due to the
minimal mechanical relaxation rate) lattice compressive deformation - the
sudden and stable jump in work-hardening of crystals). The traces of this
compression may be observed by different methods over a wide range of te-
mperatures and used for the forecast of the parameters  of  technological
interest for the new SCs.

                      I.   Introduction

   In spite of a huge number of experimental  and theoretical investiga-
tions in the field of low- (LTSCs) and high-temperature superconductors
(HTSCs), their origin is at present time indeterminate and many findings
and particularly as the total combination of them have not any exhaustive
explanation [1-15].
  So, the great interest is concerned with new hypotheses, models and laws
which can shed a light on the problem and search for the new low-, high-Tc
superconductors (SCs) and testing of the original findings with experiments.
  Here we report that room-temperature parameters of positron annihilation
                     ----------------
are linearly correlated with their critical temperature of superconducting
(SC) transition, $T_c$, in samples and with the abrupt change of compressive
strain in the series of powder samples of high-Tc $YBa_2Cu_3O_{(6+x)}$ (YBCO(123))
ceramic SCs with various deficiency of oxygen content. It is shown that the
key physical parameter for the IMSC transition is  rapid, appreciable and
steady work-hardening of crystalline lattice due to its low-temperature-
induced thermal contraction and mismatch-induced stresses between the oxygen-
enriched/depleted clusters-domains-nanoprecipitates-nanoinclusions and the
matrix.
Numerous analogous data on various LTSCs, HTSCs show that this is not parti-
cular to the YBCO(123) system but applies to the main classes of SCs.

                    II.   Experimental details

   Positron annihilation methods (PAM), such as the measurements of the half-width of angular correlation of positron-annihilation photons (ACPAP), Go/2, lifetime and Doppler broadening of annihilation radiation are successful atom-sensitive probes for studying of electronic structures, the local lattice strains, the type and concentration of defects in a great variety of materials [16-17].

The method of annihilating electron-positron pairs used in this work is based on conservation of their energy and momenta and is useful for studies of Fermi surfaces, local strains or dislocations, vacancies or micropores in solids. The ACPAP measurements on YBCO(123) ceramics were carried out using a spectrometer having a geometry of long slots. The angular discrimination of the installation was 1 mradn. The standard radio-active isotope of Na-22 with the activity of 10 mCu was used. The rate of counts in the maximum of spectra was 10 000. All the measurements were carried out in air at room temperature. The spectra were analysed with a standard program PAACFIT on the base of PC [18]. The ACPAP spectra obtained at room temperature were analysed in terms of the R- and S-parameters of the form of ACPAP curve: R is the parameter determined by the ratio of tail-end part of intensity of ACPAP (in the range of 7 mradn) to its intensity in the central part ( 0 mradn), and S is defined by the ratio of the area below the central part of ACPAP curve to the area of the whole spectrum.

The series of eight powder samples with various deficiency of oxygen content have been obtained by standard heat treatments in air environment [12-13]. One of the samples has been annealed at the highest temperature (T = 950 C) and does not show any superconducting properties (sintered sample with the absence of Tc). The Tc was measured with 4-probe method. Heat treatments changed the oxygen stoichiometry of the specimens thus changing their Tc markedly [12-13].

## III. Results and Discussion

The first remarkable result of this work is the positive linear correlations between the Tc, R parameter and the room-temperature half-width of ACPAP, Go/2, which are shown in Fig.1 (curves 1 and 3, compare with curve 1 in Fig.1 [8]). The width of the SC transition, delta Tc, and S parameter are linearly dependent (with negative slope) on the half-width of ACPAP (Fig.1, curves 2 and 4).

Table 1. The linear correlations between the critical temperature of the superconducting transition (SC), Tc, K (curve 1), the width of the SC transition, delta Tc, K (2),  and the room-temperature half-width of angular correlation of positron annihilation photons (ACPAP), Go/2, mradn, in the series of powder samples of YBCO (123) SC ceramics with various deficiency of oxygen content. The R (3) and S (4) values are the relative weights of various parts of ACPAP spectrum (their definitions are in the text).

|     | Go/2, mradn | 5.1   | 5.5   | 5.35  | 5.40  | 5.6   | 5.8   | 5.67  | 5.75  |
|-----|-------------|-------|-------|-------|-------|-------|-------|-------|-------|
| (1) | Tc,K        | -     | 54    | 58    | 63    | 71    | 79    | 80    | 80    |
| (2) | delta Tc,K  | 0     | 42    | 48    | 52    | 42    | 38    | 68    | 30    |
| (3) | R           | 0.286 | 0.359 | 0.324 | 0.325 | 0.335 | 0.364 | 0.370 | 0.360 |
| (4) | S           | 0.485 | -     | -     | 0.471 | -     | -     | 0.456 | 0.457 |

The linear increase of the room-temperature half-width of ACPAP with the Tc means that there is a strong physical parameter even at temperatures much higher than the Tc, which is the crucial one for the insulator-metal-superconductor (IMSC) transitions in samples.

Early works on LTSCs demonstrated the existence of moderate-low-temperature (T ~ 80K) forerunners of crystal lattice structural transformations (the temperature of structural transition, Tm < 25K) just before the SC transitions (Tc < 17K) in metal alloys V3Si, V3Ge [1,2] - see Fig.2, in YBCO (123) ceramic (critical oxygen concentrations for structural transi-tions, Xm (300K) > 0.31, Xm(5K) > 0.35 [12,13])- see Fig. 3b, [14]). These structural changes were the result of the accumulation of high internal stresses manifesting in lattice work-hardening, twin-domain structures of L-HTSCs [1,14], martensitic, dielectric and magnetic transformations in crystals [19, 20], etc.

Table 2. The change in lattice strain, e = delta a/a, (scale of 0.001) along the a-axis versus temperature for typical low-Tc  SCs: Nb3Sn (curve 1) and V3Si (curve 2). The cubic-tetragonal structure transformations occure at Tm = 43K (1) and 20.5K (2)) [1].

Nb3Sn (1)
| T, K | 5.1 | 15.7 | 18.5 | 37.8 | 40.8 | 42.0 | 42.5 | 43.0 | 50.0 | 77.3 |
|---|---|---|---|---|---|---|---|---|---|---|
| (1) e,(300K) | 2.06 | 2.01 | 1.65 | 1.27 | 1.06 | 1.25 | 0.89 | 0.32 | 0.13 | 0 |

Tc = 18.2K

V3Si (2)
| T, K | 4.6 | 11.3 | 15.2 | 17.2 | 18.1 | 18.8 | 19.2 | 19.6 | 20.0 | 22,0 |
|---|---|---|---|---|---|---|---|---|---|---|
| (2) e,(300K) | 0.697 | 0.721 | 0.721 | 0.699 | 0.699 | 0.636 | 0.504 | 0.466 | 0.485 | 0 |

Tc = 16.7-16.98 K

Numerous PAM data on HTSC materials indicate that positron localization in Cu-O2 planes [7] may be concerned with traps at oxygen vacancies [4] or not [8,9], defects of structure [10], grain boundaries [9], oxygen anions [11], with the increased conducting electron density [3,4] as a result of stress-field of lattice mismatch between the domains of various phases [4], domains and the matrix [6,20].

It is well known that the strong deformation of solids changes dramatically their electrical, magnetic and dielectric properties [6,20].

All the above results do not contradict each other only in terms of the strong, rapid and then steady lattice deformation induced by the lattices mismatch strains and their thermal contraction under cooling. The facts that it is needed the higher oxygen concentrations, Xm (higher mismatch-contraction internal stresses) for hard ceramics (T = 5K, Xm > 0.35 [13]) than for the softer ones (T= 300K, Xm > 0.31 only [12]) for crystal-lattice structural transitions and that e(X, 300K) < e(X, 5K) in YBCO(123) samples directly confirms this too (see Fig. 3b).

Fig.2 shows typical correlation between the change in deformation, e, and the Tc during cooling of LTSCs: Nb3Sn and V3Si (the same is valid for the HTSCs too) and clearly demonstrates that it is the sharp and appreciable (e ~ 0.001) deformation is enough for the onset of macroscopic flow, dislocation multiplication and point defects nucleation) jump in lattice compressive deformation that is the trigger-mechanism for the abrupt increase in electrical conductivity (IMSC transition) in LTSC - metal alloys and in HTSC - ceramics.

The works [5] evidence for this too, because they demonstrate the primary effect of lattice deformation during the variation in deficiency of oxygen content, and that the oxygen-vacancy order, or the linear-chain structure in the YBCO(123) ceramic is not essential for the SC transition.

The same scaling of Tc and changes in lattice strains, e, in L-HTSCs at high (300K, curves 1) and low (T < Tc, curves 1') temperatures - see Figs 3a, b - strictly corroborates this too. Note that for the same Tc (compare the curves 1,1'- 2,2' in Fig.3b) the room-temperature measurements again show much lower changes in lattice strain than the data of 5K. This fact directly evidences for the vital role of internal stresses and their minimum mechanical relaxation rates (heavy work-hardening state) at low temperatures in the IMSC and structural transitions in samples due to thermal, mismatch, high-pressure, irradiation-induced strains, etc.
The fact that the half-width of ACPAP is strictly correlated with the change in sample strain is the second remarkable finding of this work (Figs

1-3). The new law will be useful for the decoding of PAM data.

Table 3. The scaling of the changes in lattice compressive strains along the a-axis, e =delta a/a (curves 1, 1', scale of 0.001), and the critical temperatures of SC transition, Tc (2, 2'), versus the impurity content, X, in the low-Tc metal alloy Nb3Sn(1-X)Alx (a, there is no structural transformation, e-measurements at 300K; e,Tc(X) are the similar non-monotonous dependences [1]), and versus the oxygen content, X, in high-Tc YBa2Cu3O(6+X) ceramics (b, after the tetragonal-orthorhombic structural transformations (where the lattice parameters a(Xm) = b(Xm)) at Xm(300K) > 0.31 and Xm(5K) > 0.35, 1 - measurements at 300K [12], 1' - at 5K [13]).

a)

| X | 0 | ~0 | ~0.01 | ~0.02 | ~0.03 | ~0.04 | ~0.05 | ~0.06 | ~0.14 |
|---|---|---|---|---|---|---|---|---|---|
| (1) e,(300 K) | 1.54 | 1.63 | 1.68 | 1.56 | 1.39 | 1.14 | 1.088 | 1.089 | 0.0 |
| (2) Tc, K | 17.96 | 17.96 | 18.1 | 18.15 | 18.3 | 18.8 | 18.5 | 18.05 | 17.02 |

b)

| X | 0 | 0.35 | 0.45 | 0.58 | 0.64 | 0.73 | 0.78 | 0.81 | 0.84 | 0.95 |
|---|---|---|---|---|---|---|---|---|---|---|
| (1') e,(5K) | 0 | 0 | 7.7 | 9.1 | 9.99 | 11.02 | 11.75 | 12.01 | 12.34 | 13.09 |
| (2') Tc, K | - | 0 | 55.6 | 56.3 | 59.3 | 69.2 | 80.2 | 87.1 | 89 | 90.3 |

| X | 0.2 | 0.25 | 0.31 | 0.36 | 0.38 | 0.40 | 0.41 | 0.51 | 0.83 | 0.870 | 0.89 |
|---|---|---|---|---|---|---|---|---|---|---|---|
| (1) e,(300K) | 0.096 | 0.34 | 0.25 | 0.50 | 3.52 | 4.07 | 6.04 | 7.05 | 9.13 | 8.945 | 8.09 |
| (2) Tc, K | - | - | - | - | 37.2 | 43.7 | 58.0 | 51.2 | 88.9 | 84.0 | 89.8 |

Note the third remarkable finding of this work that it is the change in lattice strains that explain the non-monotonous behaviour of the Tc versus concentration of foreign atoms - so-called underdoped, optimally doped and overdoped ranges of the Tc(X) (Fig.3a), and that the lattice structural transformations are only the mechanical concomitant phenomena to the IMSC transitions like the twinning or point defect generation in lattice.

   All the results obtained can be well understood within the framework of the new universal model of electric conductivity in solids: numerous data on electroplastic, electromigration, electro-deposition/dissolution, electrical breakdown, ionic conductivity, redox, corrosion, etc. effects evidenced for the vital role of 2(3)-dimensional deformation and dislocation mechanisms in the electric and chemical properties of solids [6, 20].
This means that the voltage-current dependences in solids are identical to their stress-strain curves, for example: the so-called quasi-elastic stage in the plastic flow (the Hooke's law) is similar to the Ohm's law in current stressings of the same solids where the voltage plays the role of stress and the current is the strain of lattice; moreover, many sorts of departures from the linearity of the Ohm's law at high currents and voltages are determined by heavy plastic flow of matrix under current [6,20]. It is interesting to note here that 3 decades ago the non-standard current-voltage characteristics of the type-II LTSCs were considered to be the deformation curves of their flux-line lattices [15], which confirms the model [6,20] and coordinates with the data of the present work.
This study corroborates that it is the sharp jump in the compressive work-hardening of various lattices under cooling and matrix-precipitate mismatch strain that is the cause of IMSC transition due to the sudden drop of lattice-plastic-deformation losses for the nucleation and motion of charges in them. This effect is universal for various classes of materials in all-temperature range, but it is the steady and the lowest relaxation rates of mechanical stresses that is needed for the constant high work-hardening (the minimum mechanical losses) for the IMSC and structural transitions in solids.
   It is clear that the observation of SC transition can be realized only under the sample contraction at low temperatures (for the soft and hard LTSCs) or even at moderate temperatures (for the very hard HTSCs) and after the va-rious standard work-hardening treatments of crystals: particle irradiation with low and high energies and doses of irradiation, low-temperature-induced lattice contraction due to sample cooling and lattice mismatch between the second-phase clusters-domains-nanoinclusions (which are always present in any solid) and matrix [20], hydrostatic and axial pressures, substitution of foreign atoms, heat treatments, etc.

The second condition for the existence of IMSC transition is the sufficiently high level of free carrier density which may be determined by the lattice stress-induced coherent-incoherent properties of matrix/second-phase-precipitate interfaces [21].

It is well known that the various doses of lattice-work-hardening treatments stimulate crystal hardening or softening thus increasing or decreasing the temperatures of IMSC and structural transitions - compare the curves 1, 2 in Fig. 3a -the so-called underdoped-optimal doped-overdoped states in SCs. Recent data on the correlation of the Tc with the sample mechanical parameters of all classes of contemporary SCs confirm the universality of the model [20].

New approach to the IMSC and structural transitions in solids gives the principal way to optimize the sample-preparation procedures and the forecast of the properties of future SCs through the scaling of their mechanical and electrical (IMSC) parameters [20].

## IV. Conclusions

The first remarkable finding of this work is the linear correlation between the critical temperature of superconducting (SC) transition, $T_c$, and the room-temperature half-width of angular correlation of positron-annihilation photons (ACPAP), $G_o/2$, in the series of powder samples of high-Tc YBCO (123) ceramic SCs with various deficiency of oxygen content.
This correlation points to the existence of the fundamental physical parameter - the change in sample strain even at temperatures much higher than the insulator-metal-superconductor transition (IMSC) temperature, $T_c$, or the structural transformation temperature, $T_m$, which is the crucial one for these processes.

The second important finding of this study shows that it is the rapid substantial and then steady jump in lattice contraction (lattice work-hardening) that is the universal crucial reason for the IMSC and structural transitions under cooling and matrix-precipitate lattice mismatch strain in metal alloys, insulators, semiconductors, fullerenes, organic crystals [20], etc.

The similar increase of conductivity due to the abrupt standard macrostrain of the above crystals is a well-known phenomena [6,20] and agrees well with the above findings.

So, the fundamentally new idea for the origin of the IMSC transition is concerned with the sudden drop of plastic-deformation losses for nucleation and motion of charges [6,20].

The third important finding of this work is that it is the appreciable jump in lattice deformation under its cooling, substitution of foreign atoms, etc. and also the lowest mechanical relaxation rate in the series of low- and high-Tc SCs (metal alloys, ceramics, and so on) explain the non-monotonous behaviour of Tc versus the value of doping (predeformation, high pressures, irradiation dose, etc.) - the so-called underdoped, optimally doped and overdoped states in superconductors.

The new proposal of universal origin of SC gives the principal way to optimize the preparation treatments and the forecast of parameters of future SC materials.

## V. References

1. Testardi, L. R., Elastic behaviour and structural instability of high-temperature superconductors with A-15 structure. Physical Acoustics, 1973, vol. 10, pp. 193-296; Russian translation in [2], pp. 7-148.
2. Testardi, L. R., Structural instability of superconductors with A-15

structure, In: Superconducting compounds with beta-W structure. Addition I, Moscow, Mir, 1977, pp. 149-175 (in Russian).
3. Li, Y., Cui, L., Cao, G. et al., Positron annihilation study on the stress-field pinning mechanism in (Eu,Y)-123 superconductors. Physica C, 1999, v.314, No 1/2, pp. 55-68.
4. Cheng, G., Shang, J.,Dai, X. et al., Study of Ca-substitution on YBa2Cu4O8 superconductors by positron lifetime spectrum. Chin. J. Low Temp. Phys. 1995, v. 17, No 3, pp. 191-195.
5. Xiao, G., Cieplak,M., Gavrin,A. et al., High-temperature superconductivity in tetragonal perovskite structures: is oxygen-vacancy order important ? Phys. Rev. Lett., 1988, v. 60, No 14, pp. 1446-1449.
Gridneva, G.G., Bunina, O.A., Filip'ev, V.S. and Sakhnenko, V.P., Features of tetragonal-rhombic transition. Fiz. Nizkikh Temp., 1991, v.17, No 11/12, pp. 1552-1555.
6. Kisel, V.P., Dislocation mechanisms in electrical resistivity and superconductivity of nanostructures. Universality of dislocation mechanisms in electrical, magnetic, dielectric and mechanical properties of solids. Proc. of the Symp. on Micro- and Nanocryogenics, Aug 1-3,1999, Jyvaskyla, Finland, Univ.of Jyvaskyla, Res. Rep. 3/99, pp 48-51.
Kisel, V.P. Dislocation mechanisms explain the features of low-temperature heat conductivity in rare gas solids. Dislocation plasticity in the mobility of ions in solidified gases. The 2nd Intern. Conf. on Cryocrystals and Quantum Crystals, Polanica-Zdroj, 7-12.9.1997, Poland, Abstracts NoNo P1-13, P2-24.
7. Manuel,A.A. Positron annihilation in high-Tc superconductors. Present status. J. Phys. Condens. Matter, 1989, v.1, pp. SA (107-117).
8. Mikhailenkov, V.S., Tsapko,E.A.,Chernyashevskii, A.A.,Positron annihilation and oxygen deficiency in YBa2Cu3Ox. In: Magnetic and electronic properties of materials, Kiev, Naukova Dumka, 1989, pp. 202-208 (in Russian).
9. Zhou, X., Jiang, H., Zhang, Q. et al., Positron trapping in YBa2Cu3O(7-x) superconductors. Phys. Stat. Sol.(a), 1988, v. 109, No 2, pp. K(129-133).
10. Corbel, C., Bernede,P., Pascard, H. et al., Positron annihilation at defects in sintered high-Tc perovskite superconductors. Appl. Phys. A, 1989, v. 48, pp. 335-341.
11. Prokop'ev, V.P.,Positron annihilation and high-temperature superconductivity. Khimia Vysokikh Energii, 1990, v.24, No 3, pp. 276-280.
12. Graboi, I.E., Zubov, I.B., Ilushin, A.C. et al., Effect of nonstoichiometric oxygen on the structure and physical properties of YBa2Cu3O(7-x). Fiz. tverd. Tela, 1988, v. 30, No 11, pp. 3436-3443.
13. Cava, R.J., Hewat, A.W., Batlogg, B. et al., Structural anomalies, oxygen ordering and superconductivity in oxygen deficient Ba2YCu3Ox. Physica C, 1990, v. 165, No5/6, pp. 419-433.
14. Chen, W.M., Jiang, S.S., Guo, Y.C., Dou, S.X., Development of twins in YBa2Cu3Oy superconductor. J. Supercond., 1999, v. 12, No 2, pp. 421-426.
15. Chang, C.C., McKinnon, J.B.,and Rose-Innes,A.C., Peak effect in type II superconductors: yield point of the fluxon lattice. Phys. Stat. Sol. 1969, v.36, No 1, pp.205-209.
16. Positron Annihilation, ed. by P.C. Jain, R.M. Singru, and K.P. Goprisathan, World Scientific Publ.Co., Singapore, 1985.
17. Dlubek, G., Positron annihilation, In: Ausgewahlte Untersuchungsferfahren in der Metallkunde, Leipzig, VEB Deutscher Verlag, 1983.
18. Abdurasulov, Z. R., Arifov, P. U., Arutunov, N. Yu et al., Methods of positron diagnostics and decoding of positron annihilation spectra. Tashkent, Fan, 1985, 312 p. (in Russian).
19. Vintaikin, E.Z., Nosova, G.I, The mechanism of shape-memory-form effect in FCC-FCT alloys with martensitic transformation. Materialovedenie, 1999, No 11, pp. 2-5.                          +3
Kim, W.S., Kim, E.S., Yoon,K.H. Effects of Sm substitution on dielectric properties of Ca(1-x)Sm2x/3TiO3 ceramics at microwave frequencies. J. Am. Ceram. Soc. 1999, v. 82, No 8, pp. 2111-2115.
Shimomura, S.,Tajima,K.,Wakabayashi,N.et al., Effect of magnetic transitions and charge-ordering on crystal lattice in Nd0.5Sr0.5MnO3. J.Phys. Soc. Jap., 1999, v. 68, No 6, pp. 1943-1947.

20. Kisel, V.P. Universality of dislocation mechanisms in electrical, magnetic, dielectric and mechanical properties of solids. The VI Intern. Conf. on Materials and Mechanisms of Superconductivity and High-Tc Superconductors (M2S-HTSC-VI), Feb. 20-25,2000, Houston, Texas, USA, Abstract Book, Abstract No 4PO-80.
21. Kisel, V.P. et al., Dynamics of dislocations in InSb and GaAs crystals. Phil. Mag. A, 1993, v.67, pp. 343-360.

CAPTURES

Fig.1. The linear correlations between the critical temperature of the superconducting transition (SC), Tc,K (curve 1), the width of the SC transition, delta Tc, K (2), and the room-temperature half-width of angular correlation of positron annihilation photons (ACPAP), Go/2, mradn, in the series of powder samples of YBCO (123) SC ceramics with various deficiency of oxygen content. The R (3) and S (4) values are the relative weights of various parts of ACPAP spectrum (their definitions are in the text).

|     | Go/2, mradn | 5.1 | 5.5 | 5.35 | 5.40 | 5.6 | 5.8 | 5.67 | 5.75 |
|-----|-------------|-----|-----|------|------|-----|-----|------|------|
| (1) | $T_c$,K     | -   | 54  | 58   | 63   | 71  | 79  | 80   | 80   |
| (2) | delta $T_c$,K | 0 | 42  | 48   | 52   | 42  | 38  | 68   | 30   |
| (3) | R           | 0.286 | 0.359 | 0.324 | 0.325 | 0.335 | 0.364 | 0.370 | 0.360 |
| (4) | S           | 0.485 | -   | -    | 0.471 | -   | -   | 0.456 | 0.457 |

Fig.2. The change in lattice strain, e = delta a/a (scale of 0.001) along the a-axis vs temperature for typical low-Tc SCs: Nb3Sn (curve 1) and V3Si (curve 2). Arrows designate the appropriate Tc after the abrupt increase and then the steady lattice contraction deformations under cooling (after the cubic-tetragonal structure transformations at Tm, double arrows designate the Tm = 43K (1) and 20.5K (2)) [1].

$Nb_3Sn$ (1)

|     | T, K        | 5.1  | 15.7  | 18.5  | 37.8  | 40.8  | 42.0  | 42.5  | 43.0  | 50.0  | 77.3  |
|-----|-------------|------|-------|-------|-------|-------|-------|-------|-------|-------|-------|
|     | log T       | 0.71 | 1.196 | 1.267 | 1.578 | 1.611 | 1.623 | 1.628 | 1.634 | 1.699 | 1.888 |
| (1) | e,(300K)    | 2.06 | 2.01  | 1.65  | 1.27  | 1.06  | 1.25  | 0.89  | 0.32  | 0.13  | 0     |
|     | $T_c$ = 18.2K |    |       |       |       |       |       |       |       |       |       |

$V_3Si$ (2)

|     | T, K        | 4.6   | 11.3  | 15.2  | 17.2  | 18.1  | 18.8  | 19.2  | 19.6  | 20.00 | 22.0  |
|-----|-------------|-------|-------|-------|-------|-------|-------|-------|-------|-------|-------|
|     | log T       | 0.663 | 1.053 | 1.182 | 1.236 | 1.258 | 1.274 | 1.283 | 1.292 | 1.301 | 1.342 |
| (2) | e,(300K)    | 0.697 | 0.721 | 0.721 | 0.699 | 0.699 | 0.636 | 0.504 | 0.466 | 0.485 | 0.0   |
|     | $T_c$ = 16.7-16.98 K | | | | | | | | | | |

Fig.3. The scaling of the changes in lattice compressive strains along the a-axis, e = delta a/a (curves 1,1',scale of 0.001),and the critical temperatures of SC transition, Tc (2, 2'), versus the impurity content, X, in the low-Tc metal alloy Nb3Sn(1-X)Alx (a, there is no structural transformation, e-measurements at 300K; e, Tc(X) are the similar non-monotonous dependences [1]), and versus the oxygen content, X, in high-Tc YBa2Cu3O(6+X) ceramics (b, after the tetragonal-orthorhombic structural transformations (where the lattice parameters a(Xm) = b(Xm)) at Xm(300K) > 0.31 and Xm(5K) > 0.35, which are designated by the arrows; 1 - measurements at 300K [12], 1' - at 5K [13]).

a)

|     | X           | 0     | ~0    | ~0.01 | ~0.02 | ~0.03 | ~0.04 | 0.05  | 0.06  | 0.14  |
|-----|-------------|-------|-------|-------|-------|-------|-------|-------|-------|-------|
| (1) | e,(300K)    | 1.54  | 1.63  | 1.68  | 1.56  | 1.39  | 1.14  | 1.088 | 1.089 | 0.0   |
| 2)  | $T_c$, K    | 17.96 | 17.96 | 18.1  | 18.15 | 18.3  | 18.8  | 18.5  | 18.05 | 17.02 |

b)

| X      | 0     | 0.35 | 0.45 | 0.58 | 0.64 | 0.73  | 0.78  | 0.81  | 0.84  | 0.95  |
|--------|-------|------|------|------|------|-------|-------|-------|-------|-------|
| (1') e,(5K) | 0 | 0    | 7.7  | 9.1  | 9.99 | 11.02 | 11.75 | 12.01 | 12.34 | 13.09 |
| (2') Tc, K  | -  | 0    | 55.6 | 56.3 | 59.3 | 69.2  | 80.2  | 87.1  | 89    | 90.3  |

| X      | 0.2   | 0.25 | 0.31 | 0.36 | 0.38 | 0.40 | 0.41 | 0.51 | 0.83 | 0.870 | 0.89 |
|--------|-------|------|------|------|------|------|------|------|------|-------|------|
| (1) e,(300K) | 0.096 | 0.34 | 0.25 | 0.50 | 3.52 | 4.07 | 6.04 | 7.05 | 9.13 | 8.945 | 8.09 |
| (2) Tc, K    | -     | -    | -    | -    | 37.2 | 43.7 | 58.0 | 51.2 | 88.9 | 84.0  | 89.8 |

THE ADDITION IN PROOF:
---------------------
ABSTRACT:
EuroConf. on Physics and Applications of Multi-Junctions Superconducting
---------
Josephson Devices. Future Perspectives of Superconducting
Josephson Devices meeting, Maratea (ITALY), 1-6 July 2000, ABSTRACTS: Posters.
Int. Conf.  DPC-2001, France  (submitted):
----------

DIRECT EVIDENCE FOR THE COMPRESSIVE ORIGIN OF INSULATING-METAL-SUPERCONDUCTING
                          TRANSITION IN SOLIDS

                              Valery P. Kisel
Inst. of Solid State Physics, RAS, 142 432 Chernogolovka, Moscow district

The alternative approach to the electron-phonon interaction problem has
been suggested in recent works [1-3]. Estimates and direct observations [4]
evidence for that the reaction mechanism of photon/atom/ion/particle
insertion (extraction) into (from) the solid is the local plastic deformation
of matrix around them, so the voltage-current curves can be treated as the
stress-strain curves on the various scales of observation. For example: the
Ohm's law in current stressings of the same solids is similar to the so-called
quasi-elastic stage in plastic flow (the Hooke's law), where the voltage plays
the role of stress and the current is the strain of lattice.
Moreover, many sorts of non-linearities of the Ohm's law at high currents
and voltages, its serrated dependences, etc. on various scales of observation
are the results of sample micro- and macrodeformation up to the fracture
(electromigration or electrical breakdown, for example) [1-3].
Usually the drop in resistivity at the phase insulator to metal transition
during the concentration increase of impurity phase is not large but it is
such a change in the resistivity that has frequently been observed for
sintered materials especially when the composition is not stoichiometric.
The experimental data
allow to extrapolate this mechanism to photon interaction with electric
carriers and solids (radiation resisistance) [5].
Numerous literature data on optical and electric properties of solids
corroborate the new universal model of ion dynamic in which the losses for
ion plastic flow are the main source of sample optical properties, electrical
resistance and deformation [1-4,6].
Electroplastic and isotope effects, electromigration, electro-deposition/
/dissolution, electrical breakdown and ionic conductivity, redox and corrosion,
adsorption-desorption, diffusion, radiation-irradiation and ion-emission
effects, etc. evidenced for the vital role of 2(3)-dimensional deformation and
dislocation mechanisms in physical and chemical properties of solids [1-3].
   All the experimental results of this work do not contradict each other
only in terms of strong, rapid and then steady matrix deformation induced by
the lattice thermal contraction under cooling and mismatch strains between
matrix and impurity phase [2-8]. This stress in linear approximation is
                                    2
tau(T)= Go(T=0K) x [Ç(BTm) (1 - DTm) - T [alfa (T)-alfa (T)](1 - CT)],   (1)

where A, (B,D) < 1, C are the constants, $G_1$ and $T_m$ are the shear modulus and the melting point of matrix, $\alpha_{1,2}$ are the thermal expansion coefficients for matrix and second-phase inclusions, respectively, T is the absolute temperature [7].

It is easy to see that it is only this expression (1) for rapid substantial increase and then the steady (due to the minimal mechanical relaxation rate) lattice compressive stress (the sudden and stable jump in lattice work-hardening at low temperatures) that gives the correlations between high values of Tc, Tm, Go and appropriate jump changes in thermal expansion coefficients which are the well-known characteristics for many insulator-metal-superconductor (IMST), and structural transitions under cooling in metal alloys, insulators, semiconductors, fullerenes, organic crystals [1-3, 6], etc.

This correlation between the temperatures of IMST, Tc, and the room-temperature lattice strain changes for YBCO(123) ceramics with various oxygen content [1-3] pointed to the existence of the fundamental physical parameter - the change in sample strain even at temperatures much higher than the Tc or the structural transformation temperature, Tm, which is the crucial one for these processes. Moreover, note that it is the change in lattice strains that explain the isotope-effect [2] and the non-monotonous behaviour of Tc vs. concentration of foreign atoms - the so-called underdoped, optimally doped and overdoped ranges of Tc [1-3], and that the lattice structural transformations are only the mechanical concomitant phenomena to the IMSC transitions (Mott-transition) like the twinning or point defect generation in lattice, tweed quasi-periodical contrast [7] or microcracks [8], etc.

The intimate interconnection of conductivity and changes in magnetic, dielectric and other proprties of solids due to their abrupt standard macrodeformation is a well-known phenomena and agrees well with the new universal approach [1-3].

It is worth noting that it is the appreciable jump in lattice deformation under its cooling, substitution of foreign atoms, irrradiation of particles, etc. and also the lowest mechanical relaxation rate in the series of low- and high-Tc SCs (metal alloys, ceramics, and so on) explain the standard non-monotonous behaviour of Tc versus the value of doping (predeformation, high pressures, irradiation dose, etc.) - the so-called underdoped, optimally doped and overdoped states in superconductors [2-3]. So, the new fundamental finding for the origin of the IMSC and other transitions in solids is concerned with sudden and stable drop of plastic-deformation losses for the nucleation and dynamics of charges, changes in obstacles for the motion of dislocations, dielectric and magnetic domains, etc. [1-3].

REFERENCES

1. V.P. Kisel, The 2nd Intern. Conf. on Cryocrystals and Quantum Crystals, Polanica-Zdroj, 7-12.9.1997, Poland, Abstracts P1-13, P2-24.
2. V.P. Kisel et al., Proceed. of the Symp. on Micro- and Nanocryogenics, Aug 1-3,1999, Jyvaskyla, Finland, Univ. of Jyvaskyla, Res. Rep. 3/99, pp 48-51; Izvest. Uzbek Akad. Nauk. Ser Phys. 2000 (in press).
3. V.P. Kisel, The VI Intern.Conf. on Materials and Mechanisms of Super-conductivity and High-Tc Superconductors (M2S-HTSC-VI), Feb. 20-25,2000, Houston, Texas, USA, Abstract Book, Abstract 4PO-80.
4. A. Pundt, M. Getzlaff, M. Bode, R. Kirchheim and R. Wiesendanger, Phys. Rev. B, 61 (2000) 9964; P. Ruffieux, O. Groning, P. Schwaller, L. Schlapbach, and P. Groning, Phys. Rev. Lett., 84 (2000) 4910.
5. A. Sparenberg, G.L.J.A. Rikken, and B.A. Tiggelen, Phys. Rev. Lett.,79 (1997) 757;
A.M. Danischevsky, A.A. Kastalsky, S.M.Ryvkin, and I.D.Jaroshetsky, Zh. Eksper. Teor. Fiz., 58 (1970) 544;
L.E.Vorob'ev,D.V.Donetsky,and D.A.Firsov, Pis'ma v Zh. Eksper.Teor. Fiz., 71 (2000) 477.

D.I.Tetelbaum, V.A.Panteleev, A.Yu. Azov, M.V. Gutkin,
About the United Approach to Long-Range Effect Interpretatrion at
Irradiation of Solids by Charge Particles and Photons of Visible-Light
Region, Poverkhnost'.Rentgen., Sinkhrotron. i Neitronnye Issledov., 2000,
No 5, pp 87-89 (in Russian).
6. M. Suenaga and J.M. Galligan, Scripta Metall., 5 (1971) 829.
7. N.S. Kissel and V.P. Kisel, Mater. Sci. Engn. A, 2000 (Proc. of Int. Conf. on the Fundamentals of Plastic Deformation,"Dislocations 2000", June 19-22, 2000, Gaithersburg, USA; in press).
8. N. Sakai, F. Munakata, P. Diko, S. Takebayashi, S.-I. Yoo, and M. Murakami, Advances in Superconductivity X, Proc. of ISS'97, Oct. 27-30, 1997,Gifu, K.Osamura and I.Hirabayashi (Eds.),Springer,Tokyo 1998,p.645.

ABSTRACT    ( XXIV Int. Conf. on Physics in Semiconductors, 2-7 Aug 1998,
         Jerusalem, Israel, Abstracts, vol. 1, Tu-P166):

DISLOCATION MECHANISMS OF LATTICE-IMPURITY PHASE INTERACTION
        IN SOLIDS AT LOW AND HIGH TEMPERATURES
                     Valery P. Kisel
    Institute of Solid State Physics,142432 Chernogolovka,Moscow

The main findings of this work are the demonstration of  the
key-role of dislocation mechanisms in the  lattice deforma-
tion modes (DM) at various scales: atomic  (thermal (phonon)
conductivity, TC, heat capacity, HC), micro- and  mesoscopic
(thermal expansion  coefficient, TEC, "elastic stage",inter-
nal friction, dislocation  motion or multiplication), macro-
scopic (deformation/fracture) length  scales under different
tests in insulators (solid helium, alkali halides, etc.),me-
tals, amorphous alloys and  semiconductors at  low  and  high
temperatures,and the scaling behaviour of their stresses [3]
The local temperature-dependent (T) lattice mismatch  stres-
ses between nanoparticles of impurity phase  and  the  matrix
are the main obstacles for DM due to dislocation cross-slip,
climb, bowing  mechanisms in the effect of T  and  impurities
on the TC,HC,TEC, DM and electron (ionic) conductivity (EC):
1. The same "size effect", the surface roughness effect, the
sensitivity of TC,TEC,EC,HC and the parameters of DM to tem-
perature, work-hardening effects in solids at low/high T [1]
2. The same dependences of orientation angle anisotropy  for
TC, TEC and DM in the h.c.p. crystals [3].
3. The spread of TC, EC, TEC, HC and DM data points (curves)
usually increases with temperature or thermal resistance de-
crease, and it  disappeares at high temperatures [1].
4. The non-monotonous temperature behaviour of TC,HC,EC, TEC
is similar to the low-/moderate temperature and in the vici-
nity of the melting point (Tm) yield stress anomalies  (YSA)
in insulators,metal/disodered alloys, crystalline gases,etc.
Its parameters can be easily estimated (the  values of Tmax,
Tmin and the analytical forms for the left or right wings of
$k(T) = (T^{-7}$ to $T^{-3})$ and $k(T) = T^{-1}$, k is the TC, for example)
through the temperature dependence of thermal expansion mis-
match between precipitates and matrix in various solids [2].
5.Impurity nano-,micro-,macroprecipitates are always present
in  every so-called "solid solution" or alloy  with the ten-
dency to the miscibility gap at any concentration or  tempe-
rature. The local  stresses  due  to  the  precipitate-matrix
mismatch play the vital role  in  the  effect  of  dislocation
mechanisms on the TC, TEC, HC, EC and the yield-stress "ano-
malies" at low, ultra-low and Tm temperatures, and  this   is proved by the scaling of their deformation stresses [3].
   I   6. Within the framework of DM approach the pure metal (supe-
   I    rconductor)-insulator (semiconductor) transition in  EC is
   I    anologous to the transition of low-temperature YSA from soft
   I    to work-hardened (impure, predeformed, etc.) crystals [2].
        1. Kisel, V. P. et al., Phil.Mag.,1993,v.67A,No 2,p.343-360.
        2. Kisel, V.P., Mater.Sci.Forum,1993, v.119-121, pp 227-232.
        3. Kisel, V.P., II Int.Conf.on Cryocrystals and Quantum Cry-
        stals,Polanica-Zdroj,7-12.09.1997,Poland,Abstr.P1-13, P2-24.

     RADICAL CHANGE IN FUNDAMENTAL APPROACH TO INSULATOR-METAL-
             -SUPERCONDUCTOR TRANSITION

                    Valery P. Kisel
    Institute of Solid State Physics, 142432 Chernogolovka, Moscow
         E-Mail:   kisel@issp.ac.ru     Fax: (096) 576-41-11

It is well known that strong influences (deformation, pressures,im-
purity-phase precipitates, particle  irradiation, heat  treatments,
etc.) can noticeably change  the dislocation structure and internal
stresses of materials, or  electrical  resistivity/conductivity  of
solids like it is under electrical charge  nucleation (the same re-
gion around electron is called polaron)  and motion in lattice [1].
The deformation  origin of the above  processes  allows one to des-
cribe them well  by standard mechanisms of  dislocation motion  and
multiplication [1],so the current-voltage  plots are similar to the
stress-strain curves. The new approach  makes absolutely clear  the
fact that every strong work-hardening of  lattice considerably rai-
ses the stress-srain (current-voltage) slope,i.e. high work-harden-
ing decreases the plastic-deformation  losses for nucleation/motion
of charges thus raising the conductivity or lowering the electrical
resistivity of material. This is absolutely  evident in [2]. So the
critical current can be treated as the yield stress, and the tempe-
rature behaviour of conductivity is related to the universal origin
of temperature-yield stress anomalies in various solids [1].
1. Kisel, V.P., II Internat.Conference on Cryocrystals and Quantum
Crystals,Polanica-Zdroj,7-12.09.1997,Poland,Abstracts P1-13,P2-24.
2. Ostrovskii,I.V.,Salivonov,I.N.,Fiz.Nizkikh Temp.1998,v.24,67-70.

    NEW APPROACH TO THE UNDERSTANDING OF ISOTOPIC AND CuO2 PLANE
           BUCKLING EFFECT IN SUPERCONDUCTORS

                    Valery P. Kisel
    Institute of Solid State Physics, 142432 Chernogolovka, Moscow
        E-mail:  kisel@issp.ac.ru         Fax: +7(096) 576-41-11

Second-phase  nano-, microprecipitates are always present  in  every
so-called "solid solution",alloy or ceramic with the tendency to the miscibility gap at any concentration or temperature. The local stresses due to the coherent matrix-particle temperature-dependent lattice mismatch play the vital role in the effect of dislocation mechanisms on temperature yield-stress "anomalies",insulator-metal-superconductor transition (Tc) [1]. This approach allows one to consider the isotopic, second phase or $CuO_2$ plane buckling effects as a result of sharp increase or decrease of internal stresses in matrix from the rise or reduction of lattice parameters of precipitates (especially in hard c-direction). The loss of coherency of nano-precipitates due to the oxygen ambient or coagulation of fine particles under annealing has to decrease the Tc like in the work [2]. In the frames of the suggested deformation model of resistivity [1] the effect of magnetic field on electrical conductivity (GMR/CMR) can be co nsidered as the magnetoplastic effect due to the effect of Lorenz force on mobile charged dislocation segments under the current stress.

1. Kisel, V.P. The New3SC -2 Conference Abstract Booklet
2. Chmaissem, O. et al.,Nature (London),1999,v.397,no 6714,pp.45-48.
3. Darinskaja E.V.et al,GADEST'99, Abstracts,Sept.25-28,1999,Sweden.

# UNIVERSALITY OF DISLOCATION MECHANISMS IN ELECTRICAL, MAGNETIC SUPERCONDUCTING AND MECHANICAL PROPERTIES OF SOLIDS

Valery P. Kisel
Institute of Solid State Physics, 142432 Chernogolovka, Moscow
E-Mail: kisel@issp.ac.ru    Fax: (096) 576-41-11

Second-phase nanoprecipitates are always present in every "solid solution", alloy or ceramic [1-5].Local stresses due to the coherent matrix-particle temperature-dependent lattice mismatch play the crucial role in the effect of dislocation mechanisms on widest range temperature yield-stress "anomalies", insulator-metal-superconductor transition [1,3]. Non-monotonous wide-range temperature dependences of (anti)ferromagnetic and (anti)ferroelectric,pyroelectric and dielectric properties (which depend on the amplitude of external stress fields - mechanical, electrical or magnetic effects) can be non-contradictory explained in the frames of dislocation mechanisms. The identical with dislocations unpinning/multiplication and other processes of magnetic vortex (MV) motion in flux-line lattice [1,4,5] under the current stressing [1] and magnetic field absolutely evidences for the matrix dislocations as the carriers of MV.Due to the [1] the $Hc_1$ and $Hc_2$ are the starting and yield stresses for dislocations.

1. Kisel, V.P. The New3SC-2 Conference Abstract Booklet.
2. Chou,C-C,Chang,H-Y.et al,Jpn.J.Appl.Phys.,1998,v.37,9B,pp.5269-72.
3. Kisel, V. P., Mater. Sci. Forum, 1993, v.119-121, pp 227--232.
4. Kisel, V.P., Physica Status Solidi, 1995, v.149A, No 1, pp 61-68.
5. Ohshima,S,Koseki,T.et al,Jpn.J.Appl.Phys.,1998,v.37,4A,pp.375-378.

•